# Atomistic Model Potential for PbTiO$_3$ and PMN by Fitting First Principles Results

M. Sepliarsky[*], Z. Wu, A. Asthagiri, R. E. Cohen

Carnegie Institution of Washington,

5251 Broad Branch Rd, N.W., Washington, DC 20015 USA

**Abstract**

We have developed a shell model potential to describe PbTiO$_3$ and PbMg$_{1/3}$Nb$_{2/3}$O$_3$ (PMN) by fitting to first-principles results. At zero pressure, the model reproduces the temperature behavior of PbTiO$_3$, but with a smaller transition temperature than experimentally observed. We then fit a shell model potential for the complex PMN based on the transferability of the interatomic potentials. We find that even for ordered PMN, quenching the structure gives a non-polar state, but with local polarization (off-center ions) indicative of relaxor behavior.

[*] Also at Instituto de Fisica Rosario, CONICET-UNR, 27 de Febrero 210 bis, 2000 Rosario, Argentina

## I. Introduction

We are developing a multi-scale model for relaxor ferroelectrics such as $PbZn_{1/3}Nb_{2/3}O_3$-$PbTiO_3$ (PZN-PT) or $PbMg_{1/3}Nb_{2/3}O_3$-$PbTiO_3$ (PMN-PT) by fitting a shell model to first principles calculations. This approach has been successfully applied to some ferroelectric pervoskites [1-3]. We find it possible to develop accurate and transferable potentials for ferroelectric pervoskites. In the present work, we present the development of a model description for the complex relaxor PMN starting from the simpler ferroelectric $PbTiO_3$.

## II. $PbTiO_3$ Model

The shell model [2, 4-6] phenomenologically describes the deformation of the electronic structure of an ion due to the interactions with other atoms. In the model, each atom is described by two charged particles: a massive core and massless shell. Electronic polarization effects are captured by the dipolar moment produced by the relative core-shell displacement. We use a core linked to the shell by an anharmonic spring, $V(\boldsymbol{w}) = \frac{1}{2}c_2 \boldsymbol{w}^2 + \frac{1}{24}c_4 \boldsymbol{w}^4$, where ω is the relative core-shell displacement. There are Coulombic interactions between the core and shell of different atoms, and short-range interactions between shells. In the latter case, a Rydberg (Buckingham) potential is used for the A – B, A – O and B – O (O – O) interactions. The above coefficients and core and shell charges are parameters in the model and in all there are a total of 23 parameters for $PbTiO_3$.

The parameters of the model are adjusted using least squares to reproduce extensive DFT-LDA results for $PbTiO_3$. The data used in the fitting include phonon frequencies and eigenvectors at 300 q points along high symmetry directions [7], and the energy and atomic forces at 20 different configurations, including the behavior of the soft mode pattern displacement at various volumes and distortions, and 15 optimized configurations with cubic, tetragonal, and rhombohedral symmetry at various volumes. The model reproduces many of the features of the phonon dispersion curves described in Ref. [7], such as the rotational instability at the R point ($R_{25}$ mode) with a frequency of 89i in good agreement with the LDA value of 83i. The model also reproduces the effective charges, the behavior of the energy as function of the soft mode displacement, and the sequence of phases as a function of volume. The optimized $E(V)$ curve is shown for all three structures in Figure 1 and is in excellent agreement with LDA.

We apply the potential to simulate the finite temperature behavior of $PbTiO_3$. Molecular dynamics simulations were carried out using DL-POLY [8] with a (N,$\sigma$,T) algorithm in a 6×6×6 unit cell size system with periodic boundary conditions. The time step was set to 0.4 fs and after equilibration (10000 MD steps) results were collected for 12 ps at each temperature. Simulations using 8×8×8 cells give similar results. Our main results are shown in Fig. 2, where the lattice parameters as a function of temperature are shown from our simulations and experiment [9]. The zero pressure simulation of the system gives a tetragonal low temperature phase (a = b < c, Pz ? 0). When the temperature increases, both the tetragonal distortion and $P_z$ decrease. $P_z$ goes from a value of 54 $\mu C/cm^2$ at T = 0 K to 32 $\mu C/cm^2$ at 400 K. The transition from the tetragonal to the cubic phase (a = b = c, $P_i$ = 0) takes place when the temperature is raised over 450 K, 300 K below the experimental value. The simulated behavior of $PbTiO_3$ with temperature is in excellent qualitative agreement with experiment, and the low $T_c$ is likely a consequence of the smaller equilibrium volume of the simulation with respect to the experiment. This is due to error in the LDA, which underestimates the equilibrium volume. A negative pressure of –5 GPa gives a $T_c$ 1150 K and c/a ratio of 1.10, which overestimates the experimental data. One could adjust the pressure in between 0 and –5 GPa to yield a $T_c$ close to the experimental value, but a better approach would be to use a more accurate functional such as the WDA [10].

## III. PMN Model

We built a first-principles based model potential for PbMg$_{1/3}$Nb$_{2/3}$O$_3$ (PMN) as follows. The same interactions in PbTiO$_3$ and PMN have the same value. That is, we keep the interatomic potentials obtained previously for PbTiO$_3$ that are present in PMN, and only fit the parameters related to the new interactions: charges and core-shell interaction for Mg and Nb, and short-range interactions for Mg-O and Nb-O. By comparing to LAPW results, we found that this produced an accurate potential for PMN, which is proof of the concept.

The additional information necessary to fit the model for PMN was generated with LAPW-LDA total energy calculations. We choose cubic configurations with 1:2 order (1 Mg: 2 Nb) along two directions: [111] and [001], where each cell contains 15 atoms. Energy and forces of 20 configurations were used to adjust the unknown parameters. The calculated equilibrium lattice parameter for the 1:2 [111] configuration is 3.986 ? (5-atom cell) compared to the experimental value of 4.049 ? for a PMN crystal [11]. We find the 1:2 [111] ordered case undergoes more pronounced and less symmetric displacements than the 1:2 [001] system. For the soft mode displacement along [111] for the 1:2 [111] ordered case, the model gives good agreement with LAPW results. At this volume, the relaxed structure presents a deep instability with respect to the ideal configuration of 0.255 eV/5 atom cell, more than five times bigger than PbTiO$_3$.

Experimental results on PMN indicate that Mg$^{+2}$ and Nb$^{+5}$ are not distributed randomly, but rather have regions with 1:1 short-range order with rocksalt structure [12,13]. Two models have been proposed to explain the observed order: the space charge model [12] and the random site model [13]. In the space charge model, polar regions with 1:1 order are embedded in a Nb rich matrix to compensate for the total charge. In the random site model, there are alternating planes along the [111] direction of Nb and a random mixture of 1Mg: 1Nb. We applied the model to compare energies at $T = 0$ K of six configurations with different order in a 1080-atom (6×6×6) supercell with periodic boundary conditions. In each case, the lattice parameter was fixed at the theoretical equilibrium value, and all atoms were allowed to relax. We examined three ordered structures, two of which correspond to the ones used to fit the model, and the [001] (NCC') structure. The NCC' structure has three different planes along the [001] direction, an Nb plane followed by two planes of (1 Mg: 1 Nb) in the rocksalt structure. We also examined three disordered systems, the first has an arbitrary random distribution of Mg and Nb, and the other two represent the proposed random site and space charge model.

The three configurations with order along [111] direction: random site, 1:2 [111], and NCC' have a significantly lower energy than the others. The model predicts the NCC' configuration to be the most stable structure in agreement with recent DFT calculations [14, 15]. The small differences of energy between the configurations suggest the possible coexistence of these three types of atomic arrangement in a real crystal. We examined all the structures after a quench to 0 K and found no net polarization, but the individual cells do have a wide spectrum of polarization values indicative of relaxor behavior. The distribution of $P_z$ for the [001] NCC' configuration as a function of temperature is shown in Fig. 3. We observe multiple peaks in $P_z$ distribution at low temperatures potentially indicating short-range polarization order. This feature vanishes at higher temperature where we have a smooth distribution. We have examined the local polarization and found that this structure is anti-ferroelectric. We are currently examining the temperature and applied field affects on more experimentally relevant PMN structures such as the random site model.

## V. Conclusions

We have a model description for PbTiO$_3$ and PMN fitted to first-principles calculations. Transferability of the interatomic potential allows us to develop the model in a progressive way from simple to more complex systems.

**Acknowledgements**

This work was supported by the Office of Naval Research under contract number N000149710052. We thank Ph. Ghosez for providing the PbTiO$_3$ lattice dynamics results and R. Migoni and M. Stachiotti for helpful discussions about the model. Computations were performed on the Cray SV1 at the Geophysical Laboratory, supported by NSF EAR-9975753 and the Keck Foundation, and the Center for Piezoelectrics by Design.

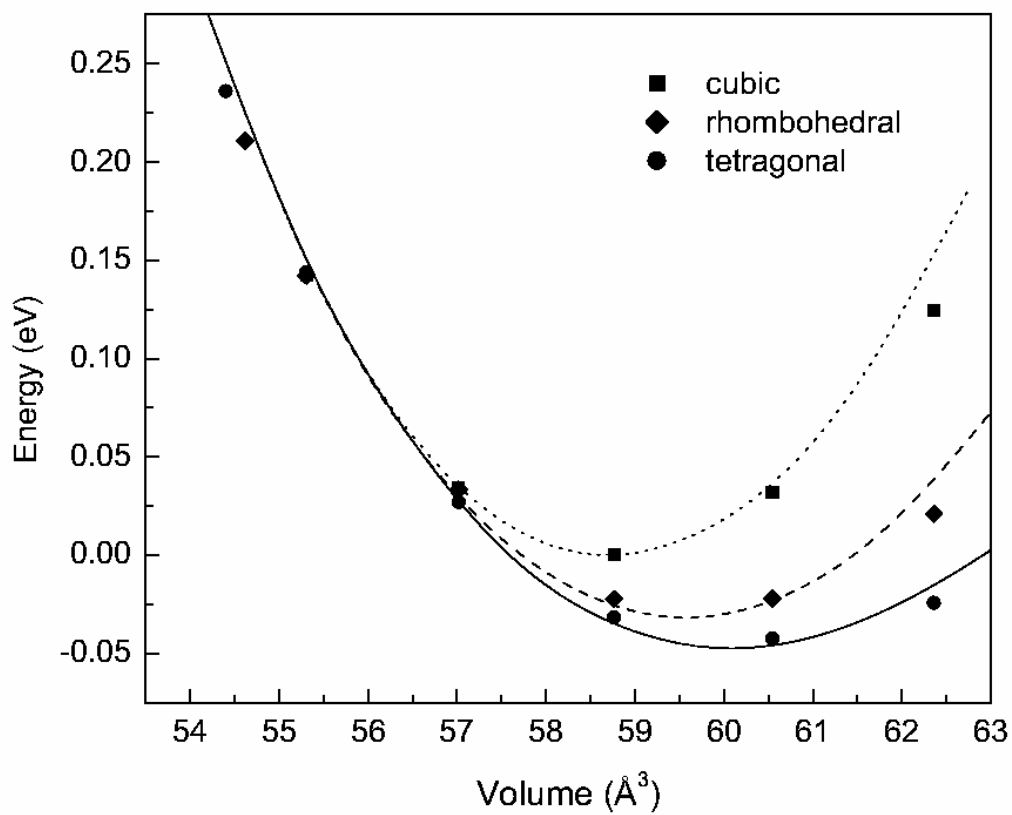

**Figure 1.** The energy versus volume curve for optimized cubic, tetragonal and rhombohedral structures using the model (lines) and LDA-DFT results (points).

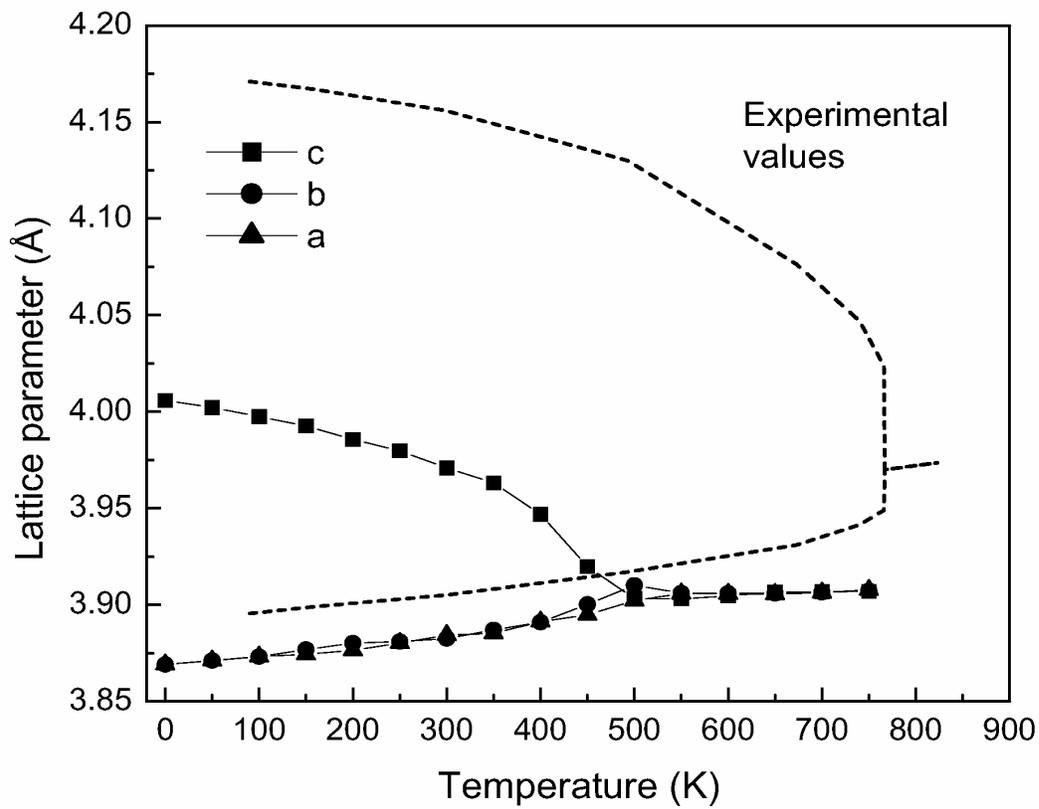

**Figure 2.** The lattice parameters for $PbTiO_3$ as a function of temperature at zero pressure obtained from MD simulations. The dashed lines are the experimental result [9].

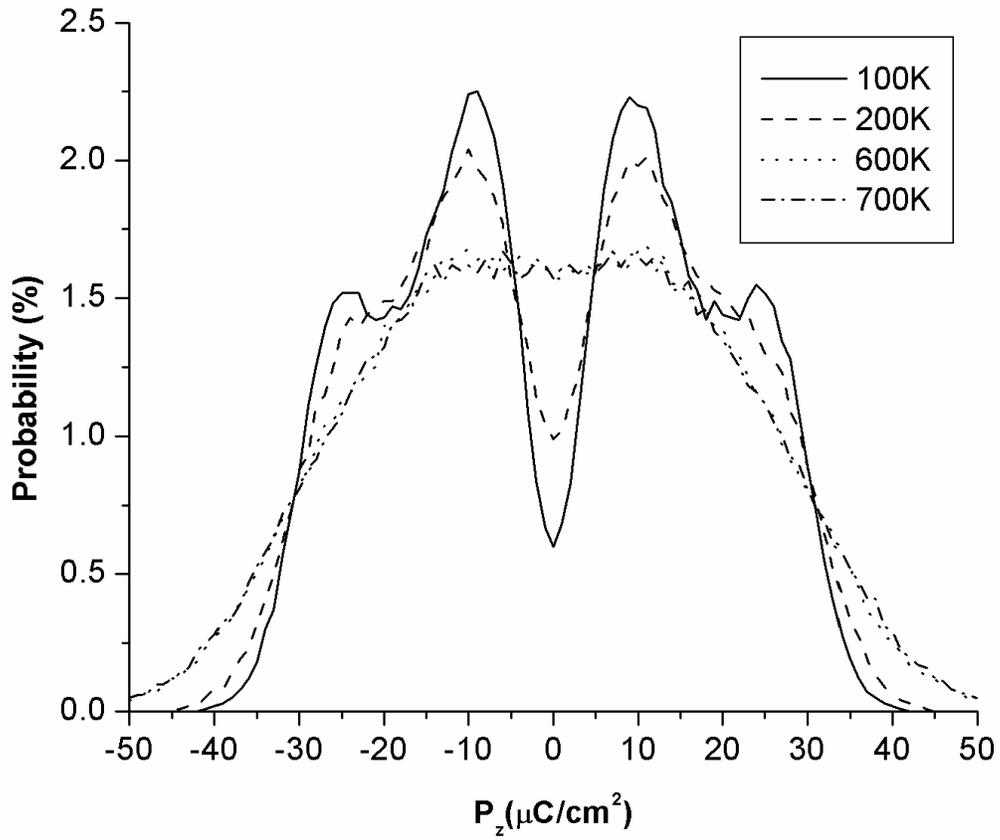

**Figure 3.** The probability distribution of the polarization in the [001] direction ($P_z$) as a function of temperature for the [001] NCC' configuration. The results were obtained from MD simulations of a 12×12×12 system with a simulation time of 12 ps.